\documentstyle[12pt]{article}
\newcommand\be{\begin{equation}}
\newcommand\ee{\end{equation}}
\newcommand\bea{\begin{eqnarray}}
\newcommand\eea{\end{eqnarray}}
\newcommand\ket[1]{|#1\rangle}
\newcommand\bra[1]{\langle #1|}

\newcommand{\fatalpha}{{\bf \alpha \kern -0.44em \alpha}}
\newcommand{\fatsigma}{{\bf \sigma \kern -0.54em \sigma}}
\newcommand{\tpchi}{{\bf \chi \kern -0.35em \chi}}
\newcommand{\llambda}{{\bf \lambda \kern -0.45em \lambda}}

\bibliography{plain}
\title{\bf Generalized qudit Choi maps  } \vspace{20mm}
\author{M. A. Jafarizadeh$^{a,b,c}$
 \thanks{E-mail:jafarizadeh@tabrizu.ac.ir}, M. Rezaee$^{a,c}$
\thanks{E-mail:karamaty@tabrizu.ac.ir}
$\;$ and S.
Ahadpour$^{a,c}$\thanks{E-mail:s.ahadpour@tabrizu.ac.ir}
.\\
\\
$^a${\small Department of Theoretical Physics and Astrophysics,
Tabriz University, Tabriz 51664, Iran.} \\ $^b${\small Institute
for Studies in Theoretical Physics and Mathematics, Tehran
19395-1795, Iran.} \\ $^c${\small Research Institute for
Fundamental Sciences, Tabriz 51664, Iran.}} \pagebreak

\begin{document}
\maketitle \vspace{15mm}
\newpage
\begin{abstract}
Following the linear programming prescription of  Ref.
\cite{PRA72}, the $d\otimes d$ Bell diagonal entanglement
witnesses are provided. By using  Jamiolkowski isomorphism, it is
shown that the corresponding positive maps are  the generalized
qudit Choi maps. Also  by manipulating particular $d\otimes d$
Bell diagonal separable states and constructing corresponding
bound entangled states, it is shown that thus obtained $d\otimes
d$ BDEW's (consequently qudit Choi maps) are non-decomposable in
certain range of their parameters.

{\bf PACs Index: 03.65.Ud }
\end{abstract}
\newpage
\vspace{70mm}
\section{Introduction}
Entanglement is one of the most fascinating features of quantum
mechanics. As Einstein, Podolsky and Rosen \cite{Einstein} pointed
out, the quantum states of two physically separated systems that
interacted in the past can defy our intuitions about the outcome
of local measurements. Moreover, it has recently been recognized
that entanglement is a very important resource in quantum
information processing\cite{nielsen}. A bipartite mixed state is
said to be separable \cite{werner} (not entangled) if considered
as a convex combination of pure product states.

One of the approaches to distinguish separable states from
entangled ones involves the so called entanglement witness (EW)
\cite{terhal}. An EW  for a given entangled state $\rho$ is an
observable W expectation value of which is nonnegative on any
separable state, but strictly negative on an entangled state
$\rho$.

There is a correspondence  relating entanglement witnesses to
linear positive (but not completely positive) maps from the
operators on Hilbert space $H_A$ to the operators on Hilbert
space  $H_B$ via Jamiolkowski isomorphism, or vice
versa\cite{jamiolkowski}.

Here in this paper, by using the prescription of Ref\cite{PRA72},
i.e., reducing the manipulation of  generic Bell-state
diagonal-entanglement witnesses to an optimization problem, we
find   $d\otimes d$ Bell states diagonal entanglement witnesses
(BDEW). By using Jamiolkowski isomorphism, we show that
corresponding positive map is the generalized qudit Choi map.
 Also  by manipulating particular qudit
Bell diagonal separable states and constructing corresponding
bound entangled states, we show that thus obtained $d\otimes d$,
BDEW's (consequently qudit Choi maps) are non-decomposable in
certain range of their parameters.

 The paper is organized as follows:\\ In section 2, we introduce the
the $d \otimes d$ Bell diagonal separable states   and
corresponding PPT states. In section 3 we give a brief review of
entanglement witness and we show that finding  generic Bell
states diagonal entanglement witnesses for $d \otimes d$ systems
reduces to a LP problem. In section 4 and 5 we find generalized
 $d\otimes d$ Choi entanglement witnesses  of first
and second types, respectively. Section 6, is devoted  to
 investigation of
non-decomposability of generalized Choi entanglement witnesss.
The paper is ended with a brief conclusion.
\section{Bell  diagonal $d \otimes d$ separable states}
Here in this section  we introduce some sets of the $d \otimes d$
Bell diagonal separable states(BDSS). In general the set of BDSS
consists of the following three categories:

{\bf Set 1:} The first set of BDSS can be constructed from Bell
states  \be
\ket{\psi_{k0}}=\frac{1}{\sqrt{d}}\sum_{l=0}^{d-1}(\omega^{k})^{l}\ket{l}\ket{l},
\ee where $\omega=e^{\frac{2\pi i}{d}}$ and $k=0,1,...,d-1$,
simply by summing over the following Bell states projection
operators:  \be
\sum_{k=0}^{d-1}\ket{\psi_{k0}}\bra{\psi_{k0}}=\frac{1}{d}\sum_{l,l^{\prime}=0}^{d-1}d\delta_{l,l^{\prime}}\ket{l}\ket{l}\otimes\bra{l^{\prime}}\bra{l^{\prime}},
\ee where we obtain one of the   BDSS  of first type,  defined as
\be\label{sep1}
\rho_{0}=\sum_{k=0}^{d-1}\ket{\psi_{k0}}\bra{\psi_{k0}}=\sum_{l=0}^{d-1}\ket{l}\bra{l}\otimes\ket{l}\bra{l}.
\ee Acting Shift operator on  (\ref{sep1}) we can find the other
BDSS of first type  as follows \be\label{sep2}
\rho_{m}=(I_{d}\otimes
S^{m})\sum_{k=0}^{d-1}\ket{\psi_{k0}}\bra{\psi_{k0}}(I_{d}\otimes
S^{m})^{\dagger}, \;\;\; m=1,2,...,d-1. \ee Therefore, the first
set of BDSS consists of the following d  BDSS \be\label{sep3}
\rho_{m}=\sum_{k=0}^{d-1}\ket{\psi_{km}}\bra{\psi_{km}}=
\sum_{l=0}^{d-1}\ket{l}\bra{l}\otimes\ket{l+m}\bra{l+m}, \;\;\;
m=0,1,...,d-1. \ee {\bf Set 2:}

In order to obtain the second set of BDSS,  we need  to consider
the following sum of projection operators
 \be
\phi^{\prime}_{k}=\sum_{i=0}^{d-1}\ket{v_{i}}\sum_{j=0}^{d-1}\ket{v_{j}}\otimes\sum_{l=0}^{d-1}\bra{v_{l}}\sum_{m=0}^{d-1}\bra{v_{m}},
\ee where $ \ket{v_{i}}=\frac{(\omega^{k})^{i}\ket{i}}{\sqrt{d}}.
$

Now   summing over the free index k and doing some routine
calculations we obtain the following  BDSS of second
type\be\label{prim1}
\rho^{\prime}_{0}=\sum_{k=0}^{d-1}\phi^{\prime}_{k}=
\sum_{k=0}^{d-1}(\sum_{i=0}^{d-1}\frac{1}{\sqrt{d}}\ket{{i}}\ket{i+k})
\otimes(\sum_{j=0}^{d-1}\frac{1}{\sqrt{d}}\bra{{j}}\bra{j+k}), \ee
where it can be written in terms of Bell states as  \be
\rho^{\prime}_{0}=\sum_{k=0}^{d-1}\ket{\psi_{0k}}\bra{\psi_{0k}}.\ee
The remaining BDSS of second type can be obtained by applying
 powers of  Modulation operator over (\ref{prim1})  separable state, \be
\rho^{\prime}_{m}=\sum_{k=0}^{d-1}(\Omega^{m}\otimes I_{d})
\ket{\psi_{0k}}\bra{\psi_{0k}}(\Omega^{m}\otimes
I_{d})^{\dagger},\;\;\; m=1,2,...,d-1.\ee Hence, the second set
of BDSS consists of the following d BDSS
 \be\label{ssep2}
\rho^{\prime}_{m}=\sum_{k=0}^{d-1}\ket{\psi_{mk}}\bra{\psi_{mk}}=\sum_{i,j,k=0}^{d-1}\omega^{m(i-j)}\ket{i}\ket{i+k}\otimes
\bra{j}\bra{j+k},\;\;\; m=0,1,...,d-1.\ee

{\bf Set 3:} Finally in order to obtain the third  set of
 BDSS, we first define the following  set of vectors  \be
\ket{u_{i}}=a_{i}{(\omega^{k})^{i}\ket{i}}, \ee \be
\ket{w_{i}}=b_{i}{(\omega^{k})^{i}\ket{i}}, \ee then using above
vectors,  we construct the following sum of projection operators
 \be
\phi^{\prime\prime}_{k}=(\sum_{i=0}^{d-1}a_{i}(\omega^{k})^{i}\ket{i})(\sum_{j=0}^{d-1}\bar{a_{j}}(\omega^{-k})^{j}\bra{j})\otimes(\sum_{i=0}^{d-1}b_{l}(\omega^{k})^{l}\ket{l})(\sum_{m=0}^{d-1}\bar{b_{m}}(\omega^{-k})^{m}\bra{m}),
\ee where we assume that $a_{i}=\bar{b_{i}}$  and
$a_{i+1}=\omega^{-ni}a_{i}$. Now, summing over free index k and
using the relation$\mid a_{i}\mid^{2}=\frac{1}{\sqrt{d}}$ and
doing some straightforward  calculations we obtain the following
BDSS's of third type \be
\rho^{\prime\prime}_{n}=\sum_{i,j,k=0}^{d-1}\omega^{nki}\frac{1}{\sqrt{d}}\ket{{i}}\ket{i+k}\otimes\frac{1}{\sqrt{d}}\bra{{j}}\bra{j+k})\omega^{-nkj},
\ee where it can be written in terms of Bell states as   \be
\rho^{\prime\prime}_{n}=\sum_{k=0}^{d-1}(\omega^{n})^{\frac{k(k+1)}{2}}\ket{\psi_{nk,k}}\bra{\psi_{nk,k}}(\omega^{-n})^{\frac{k(k+1)}{2}}.\ee
Therefore, the third  set of separable states are defined as
 \be\label{ssep3}
\rho^{\prime\prime}_{n}=\sum_{k=0}^{d-1}\ket{\psi_{nk,k}}\bra{\psi_{nk,k}},\ee
where $\rho^{\prime\prime}_{0}=\rho^{\prime}_{0}$. So far we have
introduced (3d-1) separable states. Other separable states can be
obtained by applying powers of local Shift (S) or
Modulation($\Omega$) operators on these separable states. For
instance we consider $3\otimes 3$ system which consists of  8
separable states. Now, we can obtain some other  separable states
simply by  acting powers of  Shift operator on above defined
separable states as \be\begin{array}{c} (I_{3}\otimes S^{2})
)\rho^{\prime\prime}_{1}(I_{3}\otimes
S^{2})^{\dagger}=\ket{10}\bra{10}+\ket{02}\bra{02}+\ket{21}\bra{21},\\
(I_{3}\otimes S^{2}) )\rho^{\prime\prime}_{2}(I_{3}\otimes
S^{2})^{\dagger}=\ket{20}\bra{20}+\ket{02}\bra{02}+\ket{11}\bra{11},\\(I_{3}\otimes
S) )\rho^{\prime\prime}_{1}(I_{3}\otimes
S)^{\dagger}=\ket{01}\bra{01}+\ket{12}\bra{12}+\ket{20}\bra{20},\\
(I_{3}\otimes S) )\rho^{\prime\prime}_{2}(I_{3}\otimes
S)^{\dagger}=\ket{01}\bra{01}+\ket{22}\bra{22}+\ket{10}\bra{10}.\end{array}\ee

It is easy to see that some of thus obtained BDSS's lie at the
boundary of separable states \cite{PRA72}. One can show that the
$\rho_{S}^{1}=\rho_0, \rho_{S}^{2}=\rho^{\prime}_{0}=
\rho^{\prime\prime}_{0}$ are orthogonal to the optimal $W_{red}$;
i.e., we have $Tr[W_{red}\rho_{S}^{i}]=0, i=1,2$. Hence,
$\rho_{S}^{i},i=1,2$ lie at the boundary of the separable region.
 Also convex sum of these states —i.e, $\rho_{\lambda} = \lambda\rho_{S}^{1}+(1-
\lambda)\rho_{S}^{2}$ —is orthogonal to the optimal $W_{red}$;
i.e., we have $Tr[W_{red}\rho_{\lambda}]=0$. Hence,
$\rho_{\lambda}$ lies at the boundary of the separable region.

At the end of this section we try to introduce some positive
partial transpose (PPT) operators  states which will be used
later, in the investigation of non-decomposability of generalized
Choi entanglement witnesss. We consider the following $d\otimes d$
density matrix defined as \be\label{den}
\rho_{PPT}=p(\ket{\psi_{00}}\bra{\psi_{00}})+\frac{1-p}{d}(\mu_{1}\rho_{1}+\mu_{2}\rho_{2}+...+\mu_{d-1}\rho_{d-1}),
\ee where it is positive for $0 \leq p,\mu_{1},\mu_{2},.,.,
\mu_{d-1}\leq 1$ and $\sum_{i=1}^{d-1}\mu_{i}=1$.   Now in order
to make  the partial transpose of density matrix (\ref{den}) to be
positive, i.e., to obtain PPT density matrix,  we need to use the
relations $\rho_{i}^{T_{A}}=\rho_{i}$ and substitute the partial
transpose of Bell state projection operator
$(\ket{\psi_{00}}\bra{\psi_{00}})$

\be
(\ket{\psi_{00}}\bra{\psi_{00}})^{T_{A}}=\frac{1}{d}\sum_{m,l=0}^{d-1}\omega^{ml}\ket{\psi_{m,l}}\bra{\psi_{m,d-(l)}},
\ee in partial transpose of Eq.(\ref{den}). Now, the positivity of
partial transpose of density matrix (\ref{den}) implies that
   \be\rho_{PPT}^{T_{A}}\geq 0
\Rightarrow p\leq
max\{min\{\frac{\mu_{i}}{1+\mu_{i}},\frac{\sqrt{\mu_{k}\mu_{j}}}{1+{\sqrt{\mu_{k}\mu_{j}}}}\}\}\quad\;\;,\;\;k\neq
i\neq j=1,...,d-1. \ee  For
$\mu_{1}=\mu_{2}=...=\mu_{d-1}=\frac{1}{d-1}$ the parameter p is
optimal and equal to $p=\frac{1}{d}$,  where $d\otimes d$ density
matrix reduces to  \be\label{den1}
\rho_{PPT}=[p(\ket{\psi_{00}}\bra{\psi_{00}})+\frac{1-p}{d}\sum_{i=1}^{d-1}\mu_{i}\rho_{i}],
p\leq \frac{1}{d}. \ee As we will show later in section 6, for
certain range of parameter p the density matrix (\ref{den}) become
entangled (actually bound entangled due to its PPT property). \\
In remaining part of this section we consider  particular case of
$d=3$ \be\label{den2}
\rho_{PPT}=p(\ket{\psi_{00}}\bra{\psi_{00}})+\frac{1-p}{3}(\mu_{1}\rho_{1}+\mu_{2}\rho_{2}),
\ee where the  positivity of  its partial transpose implies that
 \be p\leq
max\{\frac{\sqrt{\mu_{2}\mu_{1}}}{1+{\sqrt{\mu_{2}\mu_{1}}}}\},\ee
where for $\mu_{1}=\mu_{2}=\frac{1}{2}$  the parameter p is
optimal and it is equal to  $p=\frac{1}{3}$.

Finally we construct $3\otimes 3$  PPT state of second type by
using (\ref{ssep2}) separable states as \be\label{dden}
\rho^{\prime}_{PPT}=p(\ket{\psi_{00}}\bra{\psi_{00}})+\frac{1-p}{3}(\mu_{1}\rho^{\prime}_{1}+\mu_{2}\rho^{\prime}_{2}),
\ee where it is positive for $0 \leq p,\mu_{1},\mu_{2}\leq 1$ and
$\sum_{i=1}^{2}\mu_{i}=1$.

Now, the positivity of partial transpose of density matrix
(\ref{dden}) implies that \be{\rho^{\prime}}_{PPT}^{T_{A}}\geq 0
\Rightarrow p\leq max\{\frac{2\mu_{1}
^2-2+2\mu_{2}^2-2\mu_{1}\mu_{2}+3\mu_{2}+3\mu_{1}+3\sqrt{4+6\mu_{1}\mu_{2}-3\mu_{1}^2-3\mu_{2}^2}}{2(8+\mu_{2}^2-\mu_{1}\mu_{2}+\mu_{1}^2+3\mu_{1}+3\mu_{2})}\}.
\ee  For $\mu_{1}=\mu_{2}=\frac{1}{2}$ the parameter p is maximum
and equal to $p=\frac{1}{3}$.

\section{Bell state diagonal $d\otimes d$ entanglement witness}
In this section we give a brief outline of some of the main
features of Bell state diagonal $d\otimes d$ entanglement witness
together with linear programming prescription for finding it. For
further information  reader is referred to Ref.\cite{PRA72}.\\

 Let S be a convex compact set in a finite dimensional Banach
space. Let $\rho$ be a point in the space with $\rho\;\;\mbox
{which is not in}\;\;S$. Then there exists a hyperplane that
separates $\rho$ from S \cite{jamiolkowski,woronowicz,lec}.

A hermitian operator (an observable) W is called an entanglement
witness (EW) iff \be   \exists\rho\;\mbox{such that}\;\;
Tr(\hat{\rho}{W} ) < 0\ee \be \forall {\rho^{\prime}} \in
S\;\;\;Tr({\rho^{\prime}}\hat{W} )\geq 0.\ee

Using these definitions, we can restate the consequences of the
Hahn-Banach theorem \cite{lec} in several ways:

1- $\rho$ is entangled iff there exists a witness W such that $Tr
(\rho W) < 0$.

2- $\rho$ is a PPT entangled state iff there exists  a
non-decomposable entanglement witness W such that $Tr (\rho W) <
0$.

3- $\sigma$ is separable iff for all EW $\;\;Tr (W\sigma) \geq 0$.

From  theoretical point of view this theorem is quite powerful.
However, it is not useful to construct witnesses that detect a
given state $\rho$.

We know that  a strong relation was developed between entanglement
witnesses and positive maps\cite{jamiolkowski,woronowicz}. Notice
that an entanglement witness only gives one condition (namely $Tr
(W\rho) < 0$) while for the  map $(I_{A} \otimes \phi)\rho$ to be
positive definite, there are many conditions that have to  be
satisfied. Thus the map is much stronger, while the witnesses are
much weaker in detecting entanglement.  It is shown that this
concept is able to provide a more detailed classification of
entangled states.

Following Ref\cite{PRA72}, one can expand any trace class
observable in the Bell basis as  \be
W=\sum_{_{i_{1}i_{2}=0}}^{d-1} W_{_{i_{1}i_{2}}}
\ket{\psi_{_{i_{1}i_{2}}}}\bra{\psi_{_{i_{1}i_{2}}}} \ee where
$\ket{\psi_{_{i_{1}i_{2}}}}$ for $(0\leq i_{1}\leq d, 0\leq
i_{2}\leq d)$ stands for the orthonormal states for a $d\otimes
d$ Bell state. After  some calculations similar to those of
\cite{PRA72}, the trace-1 Bell state diagonal W observable can be
written as
  \be\label{GWE1}
W={\bf{r}}\frac{I_{d^2}}{d^2}+(1-{\bf{r}})\sum_{_{i_{1}i_{2}=0}}^{d-1}
q_{_{i_{1}i_{2}}}\ket{\psi_{_{i_{1}i_{2}}}}\bra{\psi_{_{i_{1}i_{2}}}}.
\ee The observable given by (\ref{GWE1}) is not a positive
operator and can not be an EW provided that its expectation value
on any pure product state is positive.
 For a given product state $\ket{\gamma}=\ket{\alpha}_{1}\ket{\alpha}_{2}$ the non
negativity of  \be\label{optt1} Tr(W\ket{\gamma}\bra{\gamma})\geq
0\ee implies that \be\label{p4} \frac{-d^{2}
\sum_{_{i_{1}i_{2}=0}}^{d-1}q_{_{i_{1}i_{2}}}P_{_{i_{1}i_{2}}}}{1-d^{2}\sum_{_{i_{1}i_{2}=0}}^{d-1}q_{_{i_{1}i_{2}}}P_{_{i_{1}i_{2}}}}
\leq {\bf{r}}\leq 0, \ee

where
$P_{_{i_{1}i_{2}}}=\mid<\gamma\mid\psi_{_{i_{1}i_{2}}}>\mid^{2}$.

As for the completeness of the Bell state
$\sum_{_{i_{1}i_{2}}}\ket{\psi_{_{i_{1}i_{2}}}}\bra{\psi_{_{i_{1}i_{2}}}}=1$,
the determination of ${\bf{r}}_{c}$ reduces to the following
optimization problem\cite{boyd} \be\label{cgamma}\begin{array}{cc}
\mbox{minimize} &
C_{\gamma}=\sum_{_{i_{1}i_{2}}}q_{_{i_{1}i_{2}}}P_{_{i_{1}i_{2}}}(\gamma)
\\ & 0\leq P_{_{i_{1}i_{2}}}(\gamma)\leq \frac{1}{d}\\ &
\sum_{_{i_{1}i_{2}}}P_{_{i_{1}i_{2}}}(\gamma)=1.\end{array}\ee
Always the distribution $P_{_{i_{1}i_{2}}}$ satisfies $0\leq
P_{_{i_{1}i_{2}}}(\gamma) \leq \frac{1}{d}$ for all pure product
states\cite{PRA72}. One can calculate the distributions
$P_{_{i_{1}i_{2}}}(\gamma)$, consistent with the aforementioned
optimization problem, from the information about the boundary of
feasible region. To achieve the feasible region we obtain the
extreme points corresponding to the product distributions
$P_{_{i_{1}i_{2}}}(\gamma)$ for every given product states by
applying the special conditions on $q_{_{i_{1}i_{2}}}$'s
parameters. $C_{\gamma}$ themselves are
 functions of the product distributions, and  they are  in turn are
functions of $\gamma$. They are not real variables of $\gamma$ but
the product states will be multiplicative. If this feasible region
constructs a polygon by itself, the corresponding boundary points
of the convex hull will minimize exactly  $C_{\gamma}$ in Eq.
(\ref{cgamma}). This problem is called LP , and the simplex method
is the easiest way of solving it. If the feasible region is not a
polygon, with the help of tangent planes in this region at points
which are determined either analytically or numerically one can
define new convex hull which is a polygon and has encircled  the
feasible region. The points on the boundary of the polygon can
approximately determine the minimum value $C_{\gamma}$ from
Eq.(\ref{cgamma}). Thus approximated value is obtained via LP.
\section{ Generalized qudit Choi map of first type} Compared with
qutrit Choi positive map \cite{choi} $\phi(a,b,c):M^{3}\rightarrow
M^{3}$,   Generalized qudit Choi map of first type
$\phi(a_{0},\cdots,a_{d-1}):M^{d}\rightarrow M^{d}$ is defined as

$\phi_{a_{0},\cdots,a_{d-1}}(\rho)$=$$ \left(\begin{array}{cccc}
a_{0}\rho_{11}+a_{1}\rho_{22}+\cdots+a_{d-1}\rho_{dd} & 0 & \ldots
& 0 \\0 & a_{d-1}\rho_{11}+a_{0}\rho_{22}+\cdots+a_{d-2}\rho_{dd}
& \ldots & 0\\ \vdots & \vdots & \ddots & \vdots
\\ 0 & 0 & \ldots &
a_{1}\rho_{11}+a_{2}\rho_{22}+\cdots+a_{0}\rho_{dd}\end{array}\right)$$

\be-\rho\ee
 where  $\rho\in M^{d}$. Using  Jamiolkowski
\cite{jamiolkowski} isomorphism between the positive map  and the
operators, we obtain the following $d\otimes d$ entanglement
witnesses corresponding to Choi map

$W_{Choi}=$\be\label{CCh}\frac{1}{d(a_{0}+\cdots+a_{d-1}-1)}(a_{0}\sum_{k=0}^{d-1}\ket{\psi_{k0}}\bra{\psi_{k0}}+a_{1}\sum_{k=0}^{d-1}\ket{\psi_{k1}}\bra{\psi_{k1}}+\cdots+a_{d-1}\sum_{k=0}^{d-1}\ket{\psi_{k,d-1}}\bra{\psi_{k,d-1}}-d\ket{\psi_{00}}\bra{\psi_{00}}),
\ee where it can written in terms of  separable states
(\ref{sep3}) in the following form \be\label{CCh1}
W_{Choi}=\frac{1}{d(a_{0}+\cdots+a_{d-1}-1)}(\sum_{m=0}^{d-1}a_{m}\rho_{m}-d\ket{\psi_{00}}\bra{\psi_{00}}).
\ee For product state
$\ket{\gamma}=\ket{\alpha}_{1}\ket{\alpha}_{2}=\frac{1}{\sqrt{d}}\left(\begin{array}{ccccc}1&
\omega & \omega^{2}& \ldots &
\omega^{d-1}\end{array}\right)^{T}\otimes
\frac{1}{\sqrt{d}}\left(\begin{array}{ccccc}1& \omega &
\omega^{2}& \ldots& \omega^{d-1}\end{array}\right)^{T}$ the non
negativity of (\ref{CCh1}) implies that\be\label{ww1}
a_{0}+a_{1}+...+a_{d-1}\geq d. \ee Similar to BDEW we expand
$\ket{\psi_{00}}\bra{\psi_{00}}$ using the identity operator and
the other Bell diagonal states:\be
\ket{\psi_{00}}\bra{\psi_{00}}=I_{d^{2}}-\sum_{i\ne
j=0}^{d-1}\ket{\psi_{ij}}\bra{\psi_{ij}}. \ee Then we reduce EW to
the following form $$
W_{Choi}=\frac{1}{d(a_{0}+\cdots+a_{d}-1)}(-(d-a_{0})I_{d^{2}}+d\sum_{k=1}^{d-1}\ket{\psi_{k0}}\bra{\psi_{k0}}$$
\be\label{gchoi4}
+(a_{1}+d-a_{0})\sum_{k=0}^{d-1}\ket{\psi_{k1}}\bra{\psi_{k1}}+\cdots+(a_{d-1}+d-a_{0})\sum_{k=0}^{d-1}\ket{\psi_{k,d-1}}\bra{\psi_{k,d-1}}).\ee
Comparing with BDEW (\ref{GWE1}) we have \be\label{gchoi2}
{\bf{r}}=-\frac{d(d-a_{0})}{(a_{0}+\cdots+a_{d-1}-1)}, \ee and the
EW operator is defined as
$$W_{Choi}={\bf{r}}\frac{I_{d^{2}}}{d^{2}}+\frac{(1-{\bf{r}})}{((d^{2}-1)+(1-d)a_{0}+a_{1}+\cdots+a_{d-1})}(\sum_{k=1}^{d-1}\ket{\psi_{k0}}\bra{\psi_{k0}}+(a_{1}+d-a_{0})\sum_{k=0}^{d-1}\ket{\psi_{k1}}\bra{\psi_{k1}}$$
\be\label{gchoi1}+\cdots+(a_{d-1}+d-a_{0})\sum_{k=0}^{d-1}\ket{\psi_{k,d-1}}\bra{\psi_{k,d-1}}).
\ee Note that if $\bf{r}$ is negative, as  introduced in EW above,
this operator will be  positive, but not a completely positive
map.  Using (\ref{ww1}) inequality, the minimum negative
eigenvalue of choi EW (\ref{gchoi1}) is given by \be
\frac{{\bf{r}}}{d^{2}}+(1-{\bf{r}})\frac{a_{min}+d-a_{0}}{d((d^{2}-1)+(1-d)a_{0}+a_{1}+\cdots+a_{d-1})}<
0\;\; ,\;\; a_{min}=min\{a_{0},a_{1},...,a_{d-1}\},\ee where above
inequality is satisfied for ${\bf{r}}\leq 0$ and $1 \leq a_{0}\leq
d$.

By using (\ref{optt1}) for non-negativity of the observable
$W_{choi}$ we find the distributions $P_{ij}$ as a function of
$q_{ij}$. The minimum value of $C_{\gamma}$  is obtained  from
the boundary of the feasible region, i.e., we have \be
(C_{\gamma})=\frac{1}{((d^{2}-1)+(1-d)a_{0}+a_{1}+\cdots+a_{d-1})}({\cal
P}_{1}+\frac{(a_{1}+d-a_{0})}{d}{\cal
P}_{2}+\cdots+\frac{(a_{d-1}+d-a_{0})}{d}{\cal P}_{d}), \ee where
${\cal P}_{1}=\sum_{k=1}^{d-1}P_{_{k0}},{\cal
P}_{2}=\sum_{k=0}^{d-1}P_{_{k1}}$ and ${\cal
P}_{d}=\sum_{k=0}^{d-1}P_{_{k,d-1}}$. We can find the extreme
value of $({\cal P}_{1},{\cal P}_{2},\cdots,{\cal P}_{d})$ which
is obtained under the product states
$\ket{\gamma}=\ket{\alpha}_{1}\ket{\alpha}_{2}$ as
 \be\left\{\begin{array}{c} {\cal
P}_{1}=\mid\alpha_{0}\mid^2\mid\beta_{0}\mid^2+\mid\alpha_{1}\mid^2\mid\beta_{1}\mid^2+\cdots+\mid\alpha_{d-1}\mid^2\mid\beta_{d-1}\mid^2\\
-\frac{1}{d}\mid \mid\alpha_{0}\mid
\mid\beta_{0}\mid+\mid\alpha_{1}\mid \mid\beta_{1}\mid
e^{i\phi_{1}}+\cdots+\mid\alpha_{d-1}\mid
\mid\beta_{d-1}\mid e^{i\phi_{d-1}}\mid ^2\\
{\cal
P}_{2}=\mid\alpha_{0}\mid^2\mid\beta_{1}\mid^2+\mid\alpha_{1}\mid^2\mid\beta_{2}\mid^2+\cdots+\mid\alpha_{d-1}\mid^2\mid\beta_{0}\mid^2\\
\vdots\\ {\cal
P}_{d}=\mid\alpha_{0}\mid^2\mid\beta_{d-1}\mid^2+\mid\alpha_{1}\mid^2\mid\beta_{0}\mid^2+\cdots+\mid\alpha_{d-1}\mid^2\mid\beta_{d-2}\mid^2\end{array}\right.,
\ee where $\ket{\alpha}_{1}=\left(\begin{array}{c} \alpha_{0}\\
\alpha_{1}\\ \vdots\\ \alpha_{d-1}\end{array}\right)$ and
$\ket{\alpha}_{2}=\left(\begin{array}{c} \beta_{0}\\
\beta_{1}\\ \vdots \\ \beta_{d-1}\end{array}\right)$. One can
obtain the extreme points of the $({\cal P}_{1},{\cal
P}_{2},\cdots,{\cal P}_{d})$ as \be \left\{\begin{array}{ccc}
\ket{\alpha}_{1}=\ket{\alpha}_{2}=\frac{1}{\sqrt{d}}\left(\begin{array}{c}1\\
\omega
\\ \omega^{2}\\ \vdots \\ \omega^{d-1}\end{array}\right) & \rightarrow &
({\cal P}_{1}=\frac{1}{d},{\cal P}_{2}=\frac{1}{d},\cdots,{\cal
P}_{d}=\frac{1}{d})
\\ \ket{\alpha}_{1}=\left(\begin{array}{c}1\\0\\ \vdots\\
0\end{array}\right),
\ket{\alpha}_{2}=\left(\begin{array}{c}0\\1\\
\vdots \\0\end{array}\right) & \rightarrow & ({\cal P}_{1}=0,{\cal
P}_{2}={1},{\cal P}_{3}={\cal P}_{4}=\cdots={\cal P}_{d}=0)
\\ \ket{\alpha}_{1}=\left(\begin{array}{c}1\\ 0\\ \vdots\\ 0\end{array}\right),
\ket{\alpha}_{2}=\left(\begin{array}{c}0\\
\vdots\\0\\1_{ith}\\0\\ \vdots\\0
\end{array}\right)
& \rightarrow & ({\cal P}_{1}={\cal P}_{2}=\cdots={\cal
P}_{i-1}=0,{\cal
P}_{i}=1,{\cal P}_{i+1}=\cdots={\cal P}_{d}=0) \\
\ket{\alpha}_{1}=\left(\begin{array}{c}1\\
0\\ \vdots\\ 0\end{array}\right), \ket{\alpha}_{2}=\left(\begin{array}{c}1\\
0\\ \vdots\\0 \end{array}\right) & \rightarrow & ({\cal
P}_{1}=\frac{d-1}{d},{\cal P}_{2}={\cal P}_{3}=\cdots={\cal
P}_{d}=0)\end{array}\right. \ee where $\omega=e^{\frac{2\pi
i}{d}}$.

The convex combination of all extreme points provide a convex or
a feasible region, then we have the following optimization
problem \be \left\{\begin{array}{cc} \mbox{minimize}&
(C_{\gamma})=\frac{1}{((d^{2}-1)+(1-d)a_{0}+a_{1}+\cdots+a_{d-1})}({\cal
P}_{1}+\frac{(a_{1}+d-a_{0})}{d}{\cal
P}_{2}+\cdots+\frac{(a_{d-1}+d-a_{0})}{d}{\cal P}_{d})\\
\mbox{subject to} & 1-\frac{d}{d-1}{\cal
P}_{1}-\frac{d-2}{d-1}{\cal P}_{2}-{\cal P}_{3}-\cdots-{\cal
P}_{d}\leq 0
\\& 1-\frac{d}{d-1}{\cal P}_{1}-\frac{d-2}{d-1}{\cal
P}_{3}-{\cal P}_{2}+\cdots+{\cal
P}_{d}\leq 0\\ & \vdots \\ & 1-{\cal P}_{1}-{\cal P}_{2}-\cdots-{\cal P}_{d}\geq 0  \\
& {\cal P}_{1},{\cal P}_{2},\cdots,{\cal P}_{d}\geq
0.\end{array}\right. \ee Analytically, we have been able to show
that we will have violation only from the hyperplanes
$$(d-1)-{d}{\cal P}_{1}-(d-1){\cal
P}_{i}-(d-2)\sum_{j\neq i=2}^{d}{\cal P}_{j}=0\;\;,\;\;i=2,...d.$$
Now let us assume that the maximum value of  the violation from
the planes is $\Delta<1$. Thus, the equation of the plane passing
through the new extreme points, parallel to the above plane, is
obtained. Next we derive the intersection of the following
adjacent  planes \be\left\{\begin{array}{cc}d{\cal
P}_{1}+(d-2){\cal P}_{2}+(d-1){\cal P}_{3}+\cdots+(d-1){\cal
P}_{d}-(d-1+\Delta)=0 &
\\ \vdots  &\\ d{\cal P}_{1}+(d-1){\cal P}_{2}+(d-1){\cal P}_{3}\cdots+(d-2){\cal
P}_{d}-(d-1+\Delta)=0  &\\
 {\cal P}_{1}+{\cal
P}_{2}+\cdots+{\cal P}_{d}-1=0 & \\  {\cal
P}_{i}=0\;\;\;\;\;\;\;\;\;\;\;\;\;\;\;\;,{i=1,...,d} & \\   {\cal
P}_{1}=\frac{d-1}{d}\;\;\;\;\;\;\;\;\;\;\;\;\;\;\;\;\;\;\;\;\;\;\;\;\;\;\;\;\;\;
&
\\  {\cal P}_{i}=1\;\;\;\;\;\;\;\;\;\;\;\;\;\;\;\;,{i=2,...,d} & \\
  {\cal P}_{1}+{\cal P}_{2}+\cdots+{\cal P}_{d}=\frac{d-1}{d}\;\;\;\;\;\; &
\end{array}\right. ,\ee where  new extreme
points are obtained from intersecting the above hyperplanes. Next
we calculate $C_{\gamma}$ for all the newly obtained extreme
points and compare them with one another. Some easy calculations
give the minimum value of the parameter $C_{\gamma}$ which is
independent from $\Delta$:  \be\label{cgama}
(C_{\gamma})_{{min}}=\frac{d-1}{d}\frac{1+\frac{a_{min}-a_{0}}{d}}{((d^{2}-1)+(1-d)a_{0}+a_{1}+\cdots+a_{d-1})}\;\;
,\;\;a_{min}=min\{a_{0},a_{1},...,a_{d-1}\}\ee then the critical
value of the parameter r is obtained as \be\label{rc} {\bf
r}_{c}=\frac{-d^{2}C_{\gamma_{min}}}{1-d^{2}C_{\gamma_{min}}}. \ee
For $a_{0}=a_{1}=\cdots=a_{d-1}=1$ the parameter r reduces to
$r_{c}=-d$  corresponding to the well known reduction map
\be\label{red1}
W_{red}=\sum_{i=0}^{d-1}\rho_{i}-d\ket{\psi_{_{00}}}\bra{\psi_{_{00}}}.
\ee On the other hand, EW (\ref{gchoi1}) must have positive trace
under any product state $\ket{\gamma}\bra{\gamma}$. Thus the
introduced r in (\ref{gchoi2}) must satisfy \be r\geq
r_{c}\Rightarrow -\frac{d(d-a_{0})}{(a_{0}+\cdots+a_{d-1}-1)}\geq
\frac{-d^{2}C_{\gamma_{min}}}{1-d^{2}C_{\gamma_{min}}}, \ee where
the inequality  is satisfied for all value of $1\leq a_{0}\leq
d-1$.

\section{Generalized $d
\otimes d$ Choi entanglement witness of second type} Similar to
(\ref{CCh1}) one can define the second generalized $d \otimes d$
Choi entanglement witness by second set of Bell diagonal product
states $\rho_{m}^{\prime}$ as \be
W^{\prime}_{Choi}=\sum_{i=0}^{d-1}a_{i}\rho^{\prime}_{i}-d\ket{\psi_{_{00}}}\bra{\psi_{_{00}}}.
\ee Comparing with (\ref{gchoi4}) entanglement witness
$W^{\prime}_{Choi}$ reduced to \be\label{rgchoi4}
W^{\prime}_{Choi}=\frac{1}{d(a_{0}+\cdots+a_{d-1}-1)}(-(d-a_{0})I_{d^{2}}+d\sum_{k=1}^{d-1}\ket{\psi_{0k}}\bra{\psi_{0k}}
+\sum_{k=1}^{d-1}(a_{i}+d-a_{0})\rho^{\prime}_{k}).\ee Comparing
with BDEW (\ref{GWE1}) we have \be\label{rgchoi2}
{\bf{r}}=-\frac{d(d-a_{0})}{(a_{0}+\cdots+a_{d-1}-1)}. \ee By
using (\ref{optt1}) for non-negativity of the observable
$W^{\prime}_{choi}$ we find the distributions $P_{ij}$ as a
function of $q_{ij}$. The minimum value of $C^{\prime}_{\gamma}$
is obtained  from the boundary of the feasible region, i.e., we
have \be
(C_{\gamma})=\frac{1}{((d^{2}-1)+(1-d)a_{0}+a_{1}+\cdots+a_{d-1})}({\cal
P}^{\prime}_{1}+\frac{(a_{1}+d-a_{0})}{d}{\cal
P}^{\prime}_{2}+\cdots+\frac{(a_{d-1}+d-a_{0})}{d}{\cal
P}^{\prime}_{d}), \ee where ${\cal
P}^{\prime}_{1}=\sum_{k=1}^{d-1}P_{_{0k}},{\cal
P}^{\prime}_{2}=\sum_{k=0}^{d-1}P_{_{1k}}$ and ${\cal
P}^{\prime}_{d}=\sum_{k=0}^{d-1}P_{_{d-1,k}}$. We can find the
extreme value of $({\cal P}^{\prime}_{1},{\cal
P}^{\prime}_{2},\cdots,{\cal P}^{\prime}_{d})$ which is obtained
under the product states $\ket{\gamma}=\ket{\alpha}\ket{\beta}$ as
 \be\left\{\begin{array}{c} {\cal
P}^{\prime}_{1}=\frac{1}{d}(\sum_{k,l=0}^{d-1}\alpha_{l}\beta_{l+k}(\sum_{i=0}^{d-1}\bar{\alpha}_{l+i}\bar{\beta}_{l+k+i})-\frac{1}{d}\mid
\sum_{i=0}^{d-1}\mid\alpha_{i}\mid
\mid\beta_{i}\mid e^{i\phi_{i}}\mid ^2\;\;,\;\;\phi_{0}=0\\
{\cal
P}^{\prime}_{2}=\frac{1}{d}(\sum_{k,l=0}^{d-1}\alpha_{l}\beta_{l+k}(\sum_{i=0}^{d-1}\omega^{i}\bar{\alpha}_{l+i}\bar{\beta}_{l+k+i})\\
\vdots\\ {\cal
P}^{\prime}_{d}=\frac{1}{d}(\sum_{k,l=0}^{d-1}\alpha_{l}\beta_{l+k}(\sum_{i=0}^{d-1}\omega^{(d-1)i}\bar{\alpha}_{l+i}\bar{\beta}_{l+k+i})\end{array}\right.,
\ee where $\ket{\alpha}=\left(\begin{array}{c} \alpha_{0}\\
\alpha_{1}\\ \vdots\\ \alpha_{d-1}\end{array}\right)$ and
$\ket{\beta}=\left(\begin{array}{c} \beta_{0}\\
\beta_{1}\\ \vdots \\ \beta_{d-1}\end{array}\right)$. One can
obtain the extreme points of the $({\cal P}^{\prime}_{1},{\cal
P}^{\prime}_{2},\cdots,{\cal P}^{\prime}_{d})$ as \be
\left\{\begin{array}{ccc}
\ket{\alpha}=\ket{\beta}=\left(\begin{array}{c}1\\
0
\\ 0\\ \vdots \\ 0\end{array}\right) & \rightarrow &
({\cal P}^{\prime}_{1}=\frac{1}{d},{\cal
P}^{\prime}_{2}=\frac{1}{d},\cdots,{\cal
P}^{\prime}_{d}=\frac{1}{d})
\\ \ket{\alpha}=\frac{1}{\sqrt{d}}\left(\begin{array}{c}1\\1\\ \vdots\\
1\end{array}\right),
\ket{\beta}=\frac{1}{\sqrt{d}}\left(\begin{array}{c}1\\1\\ \vdots\\
1\end{array}\right) & \rightarrow & ({\cal
P}^{\prime}_{1}=\frac{d-1}{d},{\cal P}^{\prime}_{2}=\cdots={\cal
P}^{\prime}_{d}=0)
\\ \ket{\alpha}=\frac{1}{\sqrt{d}}\left(\begin{array}{c}1\\ \omega\\ \vdots\\
\omega^{d-1}\end{array}\right),
\ket{\beta}=\frac{1}{\sqrt{d}}\left(\begin{array}{c}1\\ \vdots\\1
\end{array}\right) & \rightarrow & ({\cal P}^{\prime}_{1}=0,{\cal
P}^{\prime}_{2}=1,{\cal
P}^{\prime}_{3}=\cdots={\cal P}^{\prime}_{d}=0) \\
\ket{\alpha}=\frac{1}{\sqrt{d}}\left(\begin{array}{c}1\\
\omega^{d-1}\\ \vdots\\ {\omega^{d-1}}^{(d-1)}\end{array}\right),
\ket{\beta}=\frac{1}{\sqrt{d}}\left(\begin{array}{c}1\\
1\\ \vdots\\1 \end{array}\right) & \rightarrow & ({\cal
P}^{\prime}_{1}={\cal P}^{\prime}_{2}=\cdots={\cal
P}_{d-1}=0,{\cal P}^{\prime}_{d}=1).\end{array}\right. \ee

The convex combination of all extreme points provide a convex or a
feasible region which is completely similar to feasible region of
$W_{Choi}$ entanglement witness (\ref{CCh}) and then optimization
problem is similar to previous optimization problem and finally
$C_{\gamma_{min}}$ is  equal to (\ref{cgamma}) and critical value
for r parameter is equal to (\ref{rc}).

\section{Non-decomposibility condition for generalized Choi entanglement witnesses}
As it is explained in first section, an entanglement witness $W$
is non-decomposable  iff there  exists a bound entangled state
$\rho$ where $Tr(\rho W)<0$. First we assume that Choi's EW is
decomposable, hence it can be written as a convex sum of a
positive operator P and partial transpose of a positive operator
Q as follows \be W_{Choi}=Q^{T_{A}}+P. \ee To do so consider
(\ref{red1}) reduction EW ($W_{red}$) which is a optimal
decomposable EW (For more details see \cite{PRA72,horodecki4})
hence it  can be written  as partial transpose of positive
operator $\tilde{Q}$ i.e.,  $W_{red}=\tilde{Q}^{T_{A}}$ where
$\tilde{Q}=\sum_{i=0}^{d-1}\rho_{i}-\sum_{m,l=0}^{d-1}\omega^{ml}\ket{\psi_{_{ml}}}\ket{\psi_{_{m,d-l}}}$.
Now using the reduction map  one can  decompose the Choi's EW of
first type as   \be\label{dec2} W_{Choi}=\lambda
W_{red}+(1-\lambda)\tilde{P}, \ee where positive operator
$\tilde{P}$ is
$$\tilde{P}=(\frac{a_{0}-d}{(\sum_{i=0}^{d-1}a_{i}-1)d}+\frac{\lambda}{d})\ket{\psi_{_{00}}}\bra{\psi_{_{00}}}+(\frac{a_{0}}{(\sum_{i=0}^{d-1}a_{i}-1)d}-\frac{\lambda}{d(d-1)})\sum_{k=1}^{d-1}\ket{\psi_{_{k0}}}\bra{\psi_{_{k0}}}+$$
\be\sum_{i=1}^{d-1}(\frac{a_{i}}{(\sum_{i=0}^{d-1}a_{i}-1)d}-\frac{\lambda}{d(d-1)})\rho_{i}\geq
0. \ee The above positive operators is Bell states  diagonal,
hence its eigenvalues, i.e.,  the coefficients of Bell states
projection operators must be positive, therefore we should have
\be\label{in1}\lambda\geq
\frac{d-a_{0}}{(\sum_{i=0}^{d-1}a_{i}-1)}\ee
\be\label{in2}\lambda\leq
\frac{a_{j}(d-1)}{(\sum_{i=0}^{d-1}a_{i}-1)}\;\; ,\;\;
j=0,...,d-1.\ee Now, combining the inequalities (\ref{in1}) and
(\ref{in2}) yields \be\label{dec1} a_{i}\geq
\frac{d-a_{0}}{d-1}\;,\;i=1,...,d-1.\ee
 Also, by raining both sides of inequality (\ref{in1}) to (d-1) power and
multiplying both sides of inequality (\ref{in2}), we get
following inequality  \be
\frac{(d-a_{0})^{d-1}}{(\sum_{i=0}^{d-1}a_{i}-1)^{d-1}}\leq
\lambda^{d-1} \leq
\frac{(a_{1}a_{2}...a_{d-1})^{d-1}}{(\sum_{i=0}^{d-1}a_{i}-1)^{d-1}}
.\ee The above inequality implies that the Choi's EW is
decomposable as long as its parameters satisfy the following
inequality \be\label{dec3} a_{1}a_{2}...a_{d-1}\geq
\frac{(d-a_{0})^{d-1}}{(d-1)^{d-1}}.\ee For particular case of $3
\otimes 3$ Choi EW,
$W_{Choi}=a\rho_{0}+b\rho_{1}+c\rho_{2}-3\ket{\psi_{_{00}}}\bra{\psi_{_{00}}}$
above inequality reduces to  \be bc\geq \frac{(3-a)^{2}}{4}, \ee
which is in agreement with \cite{choi}.

If one of the inequalities (\ref{in1}) and (\ref{in2}) violates,
then operator $\tilde{P}$  will not be  positive operator any
more, consequently one cannot say that whether Choi EW is
decomposable or non-decomposable,  because decomposition
(\ref{dec2}) is not unique. In these cases, in order to find
non-decomposability conditions for $d \otimes d$ Choi's EW
(\ref{CCh1}), we try to detect  $d \otimes d$ PPT density matrix
(\ref{den1}) by them, namely we should have \be
Tr[W_{Choi}\rho_{PPT}]=(1-p)(\sum_{i=1}^{d-1}a_{i}\mu_{i})+p(a_{0}-d)<
0,\ee which yields the following lower bounds for parameter p \be
p >
\frac{\sum_{i=1}^{d-1}a_{i}\mu_{i}}{d+\sum_{i=1}^{d-1}a_{i}\mu_{i}-a_{0}}.\ee
Now, combining this lower bound with the corresponding upper one,
due to PPT property of $d \otimes d$ density matrix (\ref{den1}),
namely $p<\frac{1}{d}$, we get the following rang for parameter p
 \be
\frac{\sum_{i=1}^{d-1}a_{i}\mu_{i}}{d+\sum_{i=1}^{d-1}a_{i}\mu_{i}-a_{0}}<
p < \frac{1}{d},\ee therefore, non-decomposability condition is
\be\label{ww} (d-1)\sum_{i=1}^{d-1}a_{i}\mu_{i}< d-a_{0}. \ee
First we consider $3\otimes 3$ systems Choi EW where
non-decomposability condition (\ref{ww}) reduces to \be
b\mu_{1}+c\mu_{2} < \frac{3-a}{2}.\ee Now choosing $\mu_{1}=1,
\mu_{2}=0$  we get \be b<\frac{3-a}{2},\ee on the other hand, the
EW  (\ref{ww1}) condition implies that $c\geq\frac{3-a}{2}$.
Similarly  by choosing $\mu_{1}=0, \mu_{2}=1$  together with
(\ref{ww1}) we get \be b\geq\frac{3-a}{2}\;\;,\;\;c <
\frac{3-a}{2}. \ee Summarizing above results,  we can deduces
that Choi's witness is decomposable if its parameters satisfy
(\ref{dec3}) and it is non-decomposable otherwise. Since
violation of the inequalities (\ref{in1}) and (\ref{in2}) will be
equivalent to non-decomposability conditions provided that it
remains an EW. Since these conditions can be the same as the
non-decomposability (\ref{ww}) simply by appropriate choice of
$\mu_{i},i=1,2$.

Again one can conclude that $d\otimes d$ entanglement witness is
decomposable if its parameters satisfy (\ref{dec3}), other wise it
is non-decomposable. Discussion about the decomposability or
non-decomposability conditions of $d\otimes d$ Choi's EW is
similar to $3\otimes 3$ case. Obviously if the parameters
$a_{i},i=1,...,d-1$ satisfy condition (\ref{dec3}) it is
decomposable but if some of the inequalities (\ref{in1}) and
(\ref{in2}) violate, then one can show that, they are  equivalent
to non-decomposability condition.

Decomposability condition of   Choi's EW second type is similar to
first type since, one write
$$
W^{\prime}_{Choi}-\lambda
W_{red}=(\frac{a_{0}-d}{(\sum_{i=0}^{d-1}a_{i}-1)d}+\frac{\lambda}{d})\ket{\psi_{_{00}}}\bra{\psi_{_{00}}}+(\frac{a_{0}}{(\sum_{i=0}^{d-1}a_{i}-1)d}-\frac{\lambda}{d(d-1)})\sum_{k=1}^{d-1}\ket{\psi_{_{0k}}}\bra{\psi_{_{0k}}}$$
\be+\sum_{i=1}^{d-1}(\frac{a_{i}}{(\sum_{i=0}^{d-1}a_{i}-1)d}-\frac{\lambda}{d(d-1)})\rho^{\prime}_{i}\geq
0, \ee where the positivity of above operator yields results
similar to (\ref{dec3}).

Discussion about its non-decomposability  is hard in general.
Since it is hard to find a PPT state consisting of separable
state. So we restrict ourselves in the remaining part of this
section only to $3\otimes 3$  EW second type. Now evaluating
$Tr[W_{Choi}\rho^{\prime}_{PPT}]$ with $\rho^{\prime}_{PPT}$
given in (\ref{dden}), we get \be
Tr[W_{Choi}\rho^{\prime}_{PPT}]<0\Rightarrow
\frac{\sum_{i=1}^{2}a_{i}\mu_{i}}{3+\sum_{i=1}^{2}a_{i}\mu_{i}-a_{0}}<
p < \frac{1}{3},\ee we see that non-decomposability condition is
similar to (\ref{ww}),  therefore decomposability  and
non-decomposability conditions of EW of second type is the same
as with the first one.

\section{Conclusion}
It is  shown that finding  generalized $d\otimes d$ Choi Bell
states diagonal entanglement witnesses can be  reduced to an LP
problem. A large group of non decomposable Choi entanglement
witnesses  have been  defined by using Bell diagonal product
states and corresponding $d\otimes d$ PPT states. We hope that in
this way,  one can study the Optimality and non-decomposibility of
Choi entanglement witnesss for generic bipartite $d_{1}\otimes
d_{2}$ systems and multipartite $d_{1}\otimes d_{2}\otimes
...\otimes d_{n}$, which  are under investigation.


\begin{thebibliography}{99}
\bibitem{Einstein} A. Einstein, B. Podolsky, and N. Rosen, Phys. Rev. {\bf 47}, 777
(1935).
\bibitem{nielsen} M. N. Nielsen and I. L. Chuang, Quantum
computation and quantum information (Cambridge University Press,
Cambridge, 2000).
\bibitem{werner} R. F. Werner, Phys. Rev. A {\bf 40}, 4277 (1989).
\bibitem{terhal} B. M. Terhal, Phys. Lett. A{\bf  271}, 319
(2000).
\bibitem{jamiolkowski} A. Jamiolkowski, Rep. Mat, Phys,
{\bf 3}, 275 (1972).
\bibitem{PRA72} M. A. Jafarizadeh, M. Rezaee, S. K. S. yagoobi
Phys . Rev. {\bf 72}, 062106 (2005).
\bibitem{woronowicz} S. L. Woronowicz, Rep. on Math. Phys. {\bf  10}, 165
(1976).
\bibitem{lec} M. Lewenstein, Quantum Information Theory, Institute
for Theoretical Physics, Unversity of Hannover, March 31, (2004).
\bibitem{choi} M. D. Choi, Linear Algebra and its Applications {\bf 12}, 95 (1975).
\bibitem{boyd} S. Boyd and L. Vandenberghe,
\emph{Convex Optimization}, Cambridge University Press, (2004).
\bibitem{horodecki4} R. Horodecki and M. Horodecki, Phys. Rev. A {\bf 54}, 1838 (1996).
\end{thebibliography}
\end{document}